\documentclass[12pt]{article}
\usepackage{epsf,epsfig,float,amssymb,latexsym,amsmath,amsthm,fancyhdr}
\usepackage{graphics,longtable,cite}
\renewcommand{\L}{\mathcal{L}}
\newcommand{\vp}{\mathbf{p}}
\newcommand{\bl}{\begin{aligned}}
\newcommand{\el}{\end{aligned}}

\newcommand{\ra}{\rangle}

\newcommand{\Rea}{{\rm Re}\,}

\newcommand{\be}{\begin{equation}}
\newcommand{\ee}{\end{equation}}
\newcommand{\ba}{\begin{eqnarray}}
\newcommand{\ea}{\end{eqnarray}}
\newcommand{\bg}{\begin{align}}
\newcommand{\egg}{\end{align}}

\newcommand{\nn}{\nonumber}

\newcommand{\ep}{\varepsilon}

\begin{document}

\title{Dynamical generation of pseudoscalar resonances}

\author{M.~Albaladejo, J.~A. Oller and L.~Roca \\
Departamento de F\'{\i}sica, Universidad de Murcia,\\ E-30071 Murcia, Spain }

\maketitle

\begin{abstract}
\noindent
We study  the interactions between the $f_0(980)$ and $a_0(980)$  scalar resonances and the  
lightest pseudoscalar mesons. We first obtain the elementary interaction amplitudes, or 
interacting kernels, without including any ad hoc free parameter. This is achieved by using 
previous results on the nature of the lightest scalar resonances as dynamically generated 
from the rescattering of S-wave two-meson pairs. Afterwards, the interaction kernels are unitarized and the final S-wave amplitudes result. 
We find that these interactions are very rich and generate a large amount of pseudoscalar resonances that could be associated with the   $K(1460)$, $\pi(1300)$, $\pi(1800)$, $\eta(1475)$ and $X(1835)$.  We also consider the exotic channels with isospin 3/2 and 1, having the latter positive G-parity. The former could 
be also resonant in agreement with a previous prediction.  
\end{abstract}

\newpage
\section{Introduction}
\label{sec:intro}
\def\theequation{\arabic{section}.\arabic{equation}}
\setcounter{equation}{0}

Due to the spontaneous chiral symmetry breaking  of strong interactions  \cite{nambu,gellman,weinberg,witten} 
strong constraints among the interactions between the lightest pseudoscalars 
arise, which are most efficiently derived in the framework of Chiral Perturbation Theory (CHPT) \cite{wein,physica,gas1,gas2}. For the isospin ($I$) 0,~1 and 1/2 the scattering of the pseudoscalars in S-wave is strong enough to  generate dynamically the lightest scalar resonances, namely, the $f_0(980)$, $a_0(980)$, 
$\kappa$ and $\sigma$, as shown in refs.~\cite{npa,doba,iamprl,iamcc,nd}. 
Still one can make use of the tightly constrained interactions among the lightest pseudoscalars in order to work out approximately the scattering between the latter mesons and scalar resonances, as we show below. We concentrate here on the much narrower resonances $f_0(980)$ and $a_0(980)$ and 
consider their interactions  with the pseudoscalars $\pi$, $K$, $\eta$ and $\eta'$.  If these interactions are strong enough new pseudoscalar resonances with $J^{PC}=0^{-+}$ would come up. 
This is the case and the resulting pseudoscalar resonances have a mass larger than 1~GeV 
(this energy limit is close to the masses of the $f_0(980)$ or $a_0(980)$), typically following the relevant scalar-pseudoscalar thresholds.

The problem of the excited pseudoscalars above 1~GeV is interesting by itself. These resonances are not typically well-known  \cite{pdg}. In $I=1/2$ one has 
the $K(1460)$ and $K(1630)$ resonances.
 The $I=1$ resonances $\pi(1300)$, $\pi(1800)$ are  somewhat better known \cite{pdg}. They are broad resonances with a large uncertainty in the width 
of the former, which is reported to range between 200-600~MeV in the PDG \cite{pdg}. Some controversy exists for interpreting the decay 
channels of the $\pi(1800)$ within a quarkonium picture \cite{barnes,klempt}. 
 It was suggested in \cite{barnes} that together with the second radial excitation of the pion there would be a hybrid resonance somewhat higher in mass \cite{barnes,hydecays}. Special mention deserves the $I=0$ channel where the $\eta(1295)$, 
$\eta(1405)$, $\eta(1475)$ have been object of an intense theoretical and 
experimental study. For an exhaustive review on the experiments performed 
on these resonances and the nearby  $1^{++}$ axial-vector resonance $f_1(1420)$ see ref.~\cite{masoni_rev}. Experimentally it has been established that while the $\eta(1405)$ decays mainly to 
$a_0\pi$ the $\eta(1475)$ decays to $K^*\bar{K}+c.c$ \cite{pdg,masoni_rev}. In this way, the study of the $\eta\pi\pi$ system is certainly the most adequate one
for isolating the $\eta(1405)$ resonance because both the $f_1(1420)$ and 
$\eta(1475)$ have a suppressed partial decay width to this channel \cite{pdg}. Refs.~\cite{pdg,masoni_rev} favor the interpretation of considering the 
$\eta(1295)$ and $\eta(1475)$ as ideally mixed states (because the $\eta(1295)$ 
and the $\pi(1300)$ are close in mass) of the same nonet of pseudoscalar resonances with the other members being the $\pi(1300)$ and  $K(1460)$. All these resonances would be the first radial excitation of the lightest pseudoscalars \cite{barnes}.
 The $\eta(1405)$ would then be an extra state in this classification whose clear signal in gluon-rich process, like $p\bar{p}$ \cite{crystal,obelix} or $J/\Psi$ radiative decays \cite{mark,dm2}, and its absence in $\gamma\gamma$ collision \cite{l3}, would favor its interpretation as a 
glueball in QCD \cite{chano,close}. 
However, this interpretation opens in turn a serious problem because present results from lattice QCD 
predict the lowest mass for the pseudoscalar glueball at around 2.4~GeV \cite{bali,morn,chen}.
 Given the success of the lattice QCD prediction for the lightest scalar glueball, with a mass at around 1.7~GeV \cite{miguel,chano2}, this 
discrepancy for the pseudoscalar channel would be quite exciting. QCD sum rules \cite{ns} give a mass 
 for the lightest pseudoscalar gluonium of $2.05\pm 0.19$~GeV and an upper bound 
 of $2.34\pm 0.42$~GeV.
 However, the $\eta(1405)$ would fit as a $0^{-+}$ glueball if the latter is 
 a closed gluonic fluxtube  \cite{faddev}. 
On the other hand, it has also been pointed out that the mass and properties of the $\eta(1405)$ are consistent with predictions for a gluino-gluino bound state \cite{17,18,close}. 
 The previous whole picture for classifying the lightest pseudoscalar resonances has been challenged in ref.~\cite{klempt}. The authors question the existence of the $\eta(1295)$ and argue that, due to a node in the $^3 P_0$ wave function 
 of the $\eta(1475)$ \cite{1007}, only one isoscalar pseudoscalar resonance in the 1.4-1.5~GeV region exists. 
 This node shifts the resonant peak position depending on the channel, $a_0\pi$ or $K^*\bar{K}+c.c$. 
 It has been also recently observed by the BES Collaboration the resonance $X(1835)$ with quantum numbers favored as a pseudoscalar $0^{-+}$ resonance  
both in   $J/\Psi\to \gamma p\bar{p}$ \cite{ablikim03} and in 
$J/\Psi\to \gamma \pi^+ \pi^-\eta^{\prime}$ \cite{ablikim05}. For the former decay ref.~\cite{johan} offers an alternative explanation in terms of the $p\bar{p}$ final state interactions. 

We consider here  the S-wave interactions between 
the scalar resonances $f_0(980)$ and $a_0(980)$ with the pseudoscalar mesons $\pi$, $K$, 
$\eta$ and $\eta'$.  The approach followed is an extended version of the one that refs.~\cite{alva1,alva2} applied to study the S-wave interactions  of the $\phi(1020)$ with the $f_0(980)$ and $a_0(980)$ resonances, respectively. We show that the interactions derived generate resonances  dynamically
that can be associated with many of the previous pseudoscalar resonances listed above, namely, with the $K(1460)$, $\pi(1300)$, $\pi(1800)$, $\eta(1475)$ and 
$X(1835)$. 
In this way, new contributions to the physical resonant signals result from this novel mechanism not explored so far. In addition, we also study other exotic channels and find that the $I=3/2$ $a_0 K$ channel could also be resonant.

	After this introduction we present the formalism and derive the S-wave scattering amplitudes for scalar-pseudoscalar interactions in section \ref{sec2}.  
Section~\ref{sec3} is dedicated to present and discuss the results. Conclusions 
	are given in section~\ref{sec4}. 

\section{Formalism. Setting the model}
\label{sec2}

\begin{figure}\centering
\includegraphics[height=5cm,keepaspectratio]{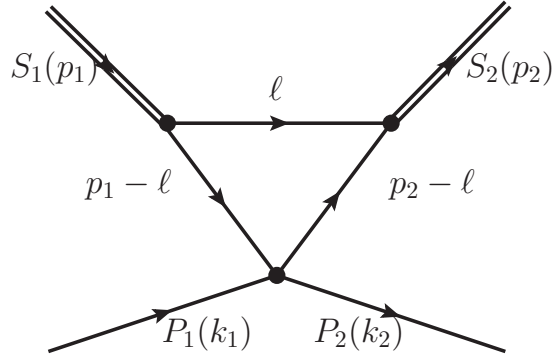}
\caption{Triangle loop for calculating the interacting kernel for 
$S_1(p_1) P_1(k_1) \to S_2(p_2) P_2(k_2)$, where between brackets the four-momentum 
for each particle is given. 
$S_{1,2}$ represents the initial, final scalar resonances and similarly for $P_{1,2}$ regarding
the pseudoscalar mesons. 
\label{fig:tri}}
\end{figure}

Our approach is based on the triangle diagram shown in Fig.~\ref{fig:tri} where an 
incident scalar resonance $S_1$ decays into a virtual $K\bar{K}$ pair.
 The filled dot in the vertex on the 
bottom of the diagram corresponds to the interaction of the incident (anti)kaon in the loop with 
the pseudoscalar $P_1$ giving rise to the  pseudoscalar $P_2$ and the same (anti)kaon. 
The out-going scalar resonance is denoted by $S_2$.  
The basic point is that this diagram is enhanced because the masses
 of  both  the $f_0(980)$ and  $a_0(980)$ resonances 
 are very close to the $K\bar{K}$ threshold. 
In this way, for scattering near the threshold of the reaction, one of the 
kaon lines in the bottom of the diagram is almost on-shell. Indeed, at threshold and in the limit  for the mass of the scalar equal to twice the kaon mass this diagram becomes infinite. 
This fact is discussed in detail in ref.~\cite{alva1} where it was already applied  for studying successfully the $\phi(1020)f_0(980)$ scattering 
and the associated $1^{--}$ $Y(2175)$ resonance. The  BABAR \cite{babar} and BELLE \cite{belle} data on $e^+e^-\to \phi(1020)f_0(980)$ were  reproduced accurately, where a strong peak for the latter resonance arises. An important conclusion of \cite{alva1} is that the $Y(2175)$ can be qualified as being a 
resonance dynamically generated due to the interactions between the $\phi(1020)$ and the $f_0(980)$ 
resonances, see also ref.~\cite{oset}.  This work was extended to $I=1$ in \cite{alva2} for studying the $\phi(1020)a_0(980)$ S-wave. There it was remarked the interest of measuring the cross sections $e^+e^-\to \phi(1020)\pi^0\eta$ because it is quite likely that an isovector companion of the $Y(2175)$ appears. In our 
present study, as well as in refs.~\cite{alva1,alva2}, one takes advantage of the fact that both the 
$f_0(980)$ and $a_0(980)$ resonances are dynamically generated by the 
meson-meson self-interactions \cite{npa,nd,mixing}. This conclusion 
is also shared with other approaches like refs.~\cite{juelich,isgur}. In this way, we can calculate the couplings of the scalar resonances considered 
to two pseudoscalars, including their relative phase.
 The coupling of the $f_0(980)$ and $a_0(980)$ resonances to a $K\bar{K}$ pair in $I=0$ and 1, respectively, is denoted by $g_{f_0}$ and $g_{a_0}$.
  These states $|K\bar{K}\ra_{I=0}$ and 
 $|K\bar{K}\ra_{I=1}$  are given by
 \begin{align}
 |K\bar{K}\ra_{I=0}&=-\frac{1}{\sqrt{2}}|K^+K^- + K^0\bar{K}^0\ra~,\nn\\
 |K\bar{K}\ra_{I=1}&=-\frac{1}{\sqrt{2}}|K^+K^- - K^0\bar{K}^0\ra~.
 \end{align}
 In this way, the $f_0(980)$ couples to $K^+K^-$($K^0\bar{K}^0$) as 
 $-\frac{1}{\sqrt{2}}$($-\frac{1}{\sqrt{2}}) g_{f_0}$ while the $a_0(980)$ couples as  $ - \frac{1}{\sqrt{2}}(\frac{1}{\sqrt{2}})g_{a_0}$.

 Let us indicate  by $P$ the total four-momentum $P=p_1+k_1=p_2+k_2$ in Fig.~\ref{fig:tri}.
  This diagram is given by $g_1 g_2 L_K$, with $g_1$ and $g_2$ the coupling of the initial and final scalar resonance to a $K\bar{K}$ pair, respectively, and 
$L_K$ is given by
 \begin{align}
 L_K&= i \int\frac{d^4\ell}{(2\pi)^4}\frac{T((P-\ell)^2)}{(\ell^2-m_K^2+i\ep)((p_1-\ell)^2-m_K^2+i\ep)
 ((p_2-\ell)^2-m_K^2+i\ep)}~.
 \label{tl.ref}
 \end{align}
  In this equation $T((P-\ell)^2)$ represents the interaction amplitude between the kaons with the external pseudoscalars. 
 Here, we employ 
the meson-meson scattering amplitudes obtained in ref.~\cite{nd} but now enlarged 
so that states with the pseudoscalar  $\eta'$ are included in the 
calculation of  $T((P-\ell)^2)$, as detailed in appendix~\ref{app:fits}. Interestingly, 
these amplitudes contain the poles corresponding to the scalar resonances $\sigma$, $\kappa$, 
$f_0(980)$, $a_0(980)$ and other poles in the region around 1.4~GeV \cite{nd}.

In order to proceed further we have to know the dependence of $T((P-\ell)^2)$ on its argument that 
includes the integration variable $\ell$. This can be done by writing the dispersion relation satisfied by $T(q^2)$ which is of the form
\begin{align}
T(q^2)&=T(s_A)+\sum_i\frac{q^2-s_A}{q^2-s_i}\frac{\hbox{Res}_i}{s_i-s_A}
+\frac{q^2-s_A}{\pi}\int_{s_{th}}^\infty ds'\frac{\hbox{Im}T(s')}{(s'-q^2)(s'-s_A)}~.
\label{dis.rel}
\end{align} 
One subtraction at $s_A$ has been taken because $T(q^2)$ is bound by a constant 
for $q^2\to\infty$, with $T(s_A)$ the subtraction constant. Typically, poles are also present deep in the $q^2$-complex plane located at $s_i$ whose residues are Res$_i$.  
These poles appear on the first Riemann sheet and are an artifact of the parameterization employed \cite{nd,pseudo}. For $q^2$ along the physical region they just give rise to soft extra contributions that could be mimic by a polynomial  of low degree in $q^2$. Inserting eq.~\eqref{dis.rel} 
into eq.~\eqref{tl.ref}, with $(P-\ell)^2=q^2$, one can write for  $L_K$
\begin{align}
L_K&=\left(T(s_A)+\sum\frac{\hbox{Res}_i}{s_i-s_A}\right)C_3
+\sum_i C_4(s_i) \hbox{Res}_i \nn\\
&-\frac{1}{\pi}\int_{s_{th}}^\infty ds' \hbox{Im}T(s') 
\left[\frac{C_3}{s'-s_A}+C_4(s') \right]~.
\label{tl.dis}
\end{align} 
Here we have introduced the three- and four-point Green functions $C_3$ and $C_4(M_4^2)$ defined by 
\begin{align}
C_3&=i\int\frac{d^4\ell}{(2\pi)^4}\frac{1}{(\ell^2-m_K^2+i\ep)((p_1-\ell)^2-m_K^2+i\ep)
 ((p_2-\ell)^2-m_K^2+i\ep)}~,\nn\\
 C_4(M_4^2)&=i\int\frac{d^4 \ell}{(2\pi)^4} \frac{1}{(\ell^2-m_K^2+i\ep)((p_1-\ell)^2-m_K^2+i\ep)
 ((p_2-\ell)^2-m_K^2+i\ep)}\nn\\
 &\times \frac{1}{((P-\ell)^2-M_4^2+i\epsilon)}~.
 \label{c3.c4.def}
\end{align}
Notice that $M_4^2$ can be real positive (when $M_4^2=s'$ in the dispersion relation), but it could 
also be negative or even complex when $M_4^2=s_i$ from the poles. One has still to perform the 
angular projection for $C_3$ and $C_4(M_4^2)$. Once this is done, eq.~\eqref{tl.dis} can 
still be used but with $C_3$ and $C_4(M_4^2)$ projected in S-wave, as we take for granted 
in the following. These functions and their 
S-wave projection are discussed in appendix~\ref{app:c3.c4.pro}. For $S_1(p_1) P_1(k_1)\to S_2(p_2) P_2(k_2)$ we have the usual Mandelstam variables 
$s=(p_1+k_1)^2=(p_2+k_2)^2$, $t=(p_1-p_2)^2=(k_1-k_2)^2$ and $u=(p_1-k_2)^2=(p_2-k_1)^2=
M_{S_1}^2+M_{S_2}^2+M_{P_1}^2+M_{P_2}^2-s-t$, with the masses of the particles indicated by $M$ with the subscript distinguishing between them. The dependence on the relative angle 
$\theta$  enters in $t$ as 
$t=(p_1^0-k_1^0)^2-(\vp-\vp')^2=(p_1^0-k_1^0)^2-\vp^2-\vp'^2+2|\vp||\vp'|\cos\theta$ with $\vp$ and 
$\vp'$ the CM three-momentum of the initial and final particles, respectively. 
 
 Eq.~\eqref{tl.dis} is our basic equation for evaluating the interaction kernels. One has 
 only to specify the pseudoscalars actually involved in the amplitude $T(q^2)$ according to the 
 specific reaction under consideration.  We now list all the channels involved for the different quantum numbers and indicate the actual pseudoscalar-pseudoscalar amplitudes required as the argument of $L_K$:
 
 \begin{itemize}
\item{$I=0$, $G=+1$}
	\begin{align}
T_L(a_0\pi \to a_0\pi)&=\frac{2 g_{a_0}^2}{3}L_K[4 \,T_{\pi K\to \pi K}^{I=3/2}
-T_{\pi K\to \pi K}^{I=1/2}]~,\nn\\
T_L(a_0\pi \to f_0\eta)&=2 g_{f_0}g_{a_0}L_K[T_{\eta K\to \pi K}^{I=1/2}]~,\nn\\
T_L(f_0\eta\to f_0\eta)&=2g_{f_0}^2 L_K[T_{\eta K \to \eta K}^{I=1/2}]~,\nn\\
T_L(a_0\pi\to f_0\eta')&=2 g_{f_0}g_{a_0} L_K[T_{\eta' K\to \pi K}^{I=1/2}]~,\nn\\
T_L(f_0\eta \to f_0\eta')&=2 g_{f_0}^2 L_K[T_{\eta K\to \eta' K}^{I=1/2}]~,\nn\\
T_L(f_0\eta'\to f_0\eta')&=2 g_{f_0}^2 L_K[T_{\eta' K\to \eta' K}^{I=1/2}]~.
	\label{n.i0}
	\end{align}

\item{$I=1/2$}
 \begin{align}
T_L(f_0 K\to f_0 K)&=\frac{g_{f_0}^2}{2} L_K[3 \,T_{K\bar{K}\to K\bar{K}}^{I=1} +T_{K\bar{K}\to K\bar{K}}^{I=0}]~, \nn\\
T_L(f_0K \to a_0 K)&=\frac{\sqrt{3} g_{f_0}g_{a_0}}{2} L_K[T_{K\bar{K}\to K\bar{K}}^{I=1}  - T_{K\bar{K}\to K\bar{K}}^{I=0}]~,\nn\\
T_L(a_0 K\to a_0 K)&=\frac{g_{a_0}^2}{2}L_K[3\, T_{K\bar{K}\to K\bar{K}}^{I=0}+ T_{K\bar{K}\to K\bar{K}}^{I=1}]~.
 \label{n.i12}
 \end{align}

\item{$I=1$, $G=-1$}
	\begin{align}
T_L(f_0\pi\to f_0\pi)&=\frac{2 g_{f_0}^2}{3}L_K[2\,T_{\pi K\to \pi K}^{I=3/2}
+T_{\pi K\to \pi K}^{I=1/2}]~,\nn\\
T_L(f_0\pi\to a_0\eta)&=\frac{2 g_{f_0} g_{a_0}}{\sqrt{3}}L_K[T_{\pi K\to \eta K }^{1/2}]~,\nn\\
T_L(a_0\eta\to a_0\eta)&=2 g_{a_0}^2 L_K[T_{\eta K\to \eta K}^{I=1/2}]~,\nn\\
T_L(f_0\pi\to a_0\eta')&=\frac{2 g_{f_0}g_{a_0}}{\sqrt{3}}L_K[T_{\pi K\to \eta' K}^{I=1/2}]~,\nn\\
T_L(a_0\eta\to a_0\eta')&=2 g_{a_0}^2L_K[T_{\eta K\to \eta' K}^{I=1/2}]~,\nn\\
T_L(a_0\eta'\to a_0\eta')&=2 g_{a_0}^2 L_K[T_{\eta' K\to \eta' K}^{I=1/2}]~.
	\label{n.i1.gm1}
	\end{align}

\item{$I=1$, $G=+1$}
	\begin{align}
T_L(a_0\pi\to a_0\pi)&=\frac{2 g_{a_0}^2}{3}L_K[4T_{\pi K\to \pi K}^{I=1/2}-T_{\pi K\to \pi K}^{I=3/2}]~.
	\label{n.i1.gp1}
	\end{align}

\item{$I=3/2$}
	\begin{align}
	T_L(a_0 K\to a_0 K)&=2g_{a_0}^2 L_K[T_{K\bar{K}\to K\bar{K}}^{I=1}]~.
\label{n.i32}
	\end{align}
 \end{itemize}

In the previous equations the different scalar-pseudoscalar states are pure isospin ones corresponding to the isospin $I$  indicated  for each item. This also applies to the 
pseudoscalar-pseudoscalar states, with $I$ as indicated in the 
 superscript of $T$. The symbol $G$ refers to G-parity. On the other hand the $I=3/2$ $\pi K$ amplitude, being much smaller than the $I=1/2$ one, has negligible effects, although it has been kept in the previous expressions.

\begin{figure}[h]\centering
\includegraphics[width=.95\textwidth,angle=0]{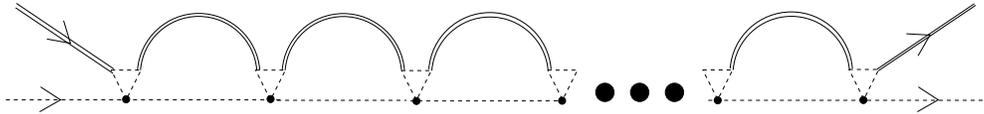}
\caption{Iteration of the interaction kernels (denoted by the triangles in the figure) by 
inserting scalar(double lines)-pseudoscalar(dashed lines) intermediate states. 
\label{fig:uni.res}}
\end{figure}

For each set of quantum numbers specified by the isospin $I$ and G-parity $G$ (if the latter is not defined this label should be omitted) we join in a symmetric matrix ${\cal T}_{IG}$ the different 
 $T_L(i\to j)$ calculated above. Then, in order to resum the unitarity loops, as indicated 
 in Fig.~\ref{fig:uni.res}, and obtain  the final  S-wave scalar-pseudoscalar T-matrix, $T_{IG}$, 
 we make use of the equation
 \begin{align}
T_{IG}&=\left[I+{\cal T}_{IG}\cdot g_{IG}(s)\right]^{-1}\cdot {\cal T}_{IG}~.
\label{t.uni}
\end{align}
For a general derivation of this equation, based on the N/D method \cite{chew}, see refs.~\cite{nd,plb} and ref.~\cite{npa}, where it is connected with the Bethe-Salpeter 
equation. In eq.~\eqref{t.uni} $g_{IG}(s)$ is a diagonal matrix whose elements are 
the scalar unitarity loop function with a scalar-pseudoscalar intermediate state. 
For the calculation of $g_{IG}(s)_i$, corresponding to  the $i_{th}$ 
state with the quantum numbers $IG$ and made up by  the scalar resonance 
$S_i$ and the pseudoscalar $P_i$, we make use of a once subtracted dispersion relation \cite{nd}. The result is
\begin{align}
g_{IG}(s)_i&=\frac{1}{(4\pi)^2}\left\{a_1+\log\frac{M_{S_i}^2}{\mu^2} 
-\frac{M_{P_i}^2-M_{S_i}^2+s}{2 s}\log\frac{M_{S_i}^2}{M_{P_i}^2} \right.\nn\\
&\left.+\frac{|\vp|}{\sqrt{s}} \Biggl[
\log(s-\Delta+2\sqrt{s}|\vp|)+\log(s+\Delta+2\sqrt{s}|\vp|)\right.\nn\\
&\left.-\log(-s+\Delta+2\sqrt{s}|\vp|)
-\log(-s-\Delta+2\sqrt{s}|\vp|)
\Biggr]\right\}
\label{g.func}
\end{align}
with $|\vp|$ the three-momentum of the channel $S_i P_i$ for a given $s$ and 
$\Delta=M_{P_i}^2-M_{S_i}^2$. The subtraction $a_1$ 
is restricted to have natural values so that the unitarity scale \cite{alva1} $4\pi f_\pi/\sqrt{|a_1|}$
 becomes not too small (e.g. below the $\rho$-mass) so that $|a_1|\lesssim 3$. In addition,
 we require the sign of $a_1$ to be negative so that resonances could be generated when the interaction kernel is positive (attractive).  

As already indicated in ref.~\cite{alva2} to ensure a continuous limit to zero $a_0(980)$ width, one has to evaluate ${\cal T}_{IG}$ at the $a_0(980)$ pole position with positive imaginary part so that $p_{1,2}^2 \to \Rea[M_{a_0}]^2 + i \epsilon$, in agreement with Eq.~\eqref{tl.ref}. Instead, in $g_{IG}(s)_{a_0 P}$, with $P$ one of the lightest pseudoscalars,  $M_{a_0}$ should appear with a negative imaginary part to guarantee that, in the zero-width limit, the sign of the imaginary part is the same as dictated by the $-i \epsilon$ prescription for masses squared of the intermediate 
states.  Such analytical extrapolations in the masses of external particles are discussed in Refs.~\cite{barton,bar1,bar2}. The same applies of course to the case of the $f_0(980)$ resonance. 

\section{Results}
\label{sec3}

In this section we show the results that follow by applying eq.~\eqref{t.uni} to the different 
channels characterized by the quantum numbers $IG$, as given in the list from eq.~\eqref{n.i0} to 
eq.~\eqref{n.i32}. 
 As discussed after  eq.~\eqref{g.func} we consider values for the subtraction constant such that
 they are negative and not very large in modulus ($|a_1|\lesssim 3$). In this way, the resonances generated 
might be qualified as dynamically generated due to the iteration of the unitarity loops. The pole positions and 
couplings of the $f_0(980)$ and $a_0(980)$ resonances are given in Tables~\ref{tab:fit_poles_1} and \ref{tab:fit_poles_2}, respectively, and they correspond to those obtained 
in the meson-meson S-wave amplitudes used, see appendix~\ref{app:fits}. 
We present the results for each of the channels with definite $IG$ separately.

\begin{figure}[h]
\includegraphics[width=\textwidth,keepaspectratio]{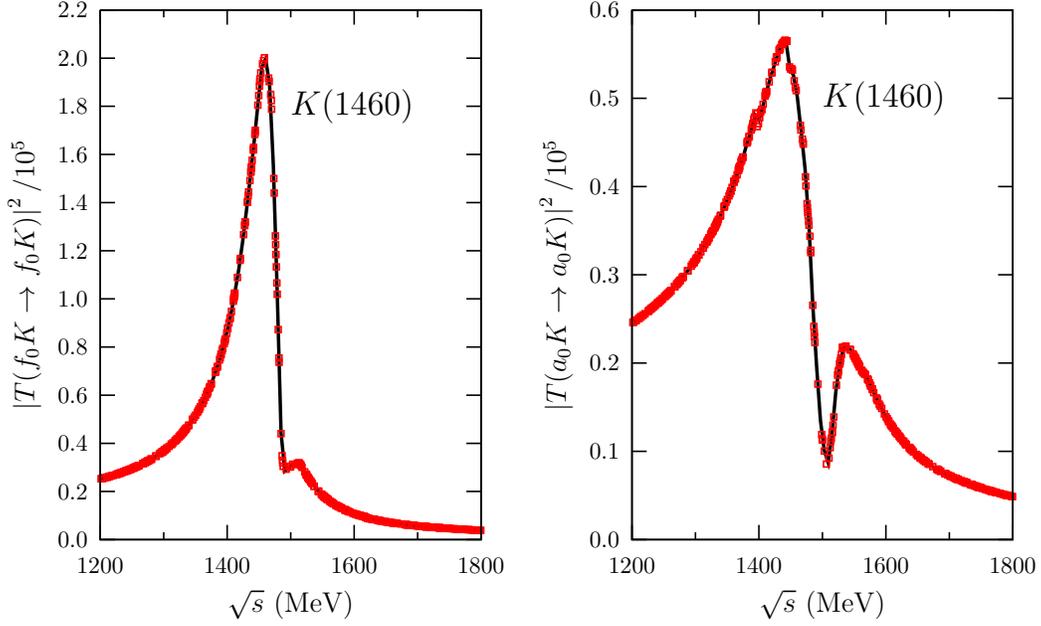}
\caption{Modulus squared of the $f_0 K\to f_0 K$ (left)  and $a_0 K\to a_0 K$ (right) S-wave amplitudes for $a_1=-0.5$~. The points correspond 
to the energies where the amplitudes have been actually calculated.
\label{fig:t2.k12}}
\end{figure}

\subsection{$I=1/2$}
First we show the results for the $I=1/2$ sector that couples together the  channels $f_0(980) K$ and $a_0(980) K$. 
 We show the modulus 
squared of the $f_0 K\to f_0 K$ and $a_0 K\to a_0 K$ S-wave amplitudes in the left and right panel of Fig.~\ref{fig:t2.k12}, 
respectively. We obtain a clear resonant peak with its maximum at $1460$~MeV for $a_1$ around $-0.5$, that 
corresponds to the nominal mass of the $K(1460)$ resonance \cite{pdg}. The results are not 
very sensitive to the actual value of $a_1$ but the position of the peak displaces to lower values for decreasing 
$a_1$ and the width somewhat increases.   The  visual width of the peak is around 100~MeV, although it appears wider in $a_0 K\to a_0 K$ scattering. 
In refs.~\cite{daum,branden} a larger width of around 250~MeV is referred. One has to take into account that the channel $K^*(892)\pi$ is not 
included and it seems  to couple strongly with the $K(1460)$ resonance \cite{pdg}.  It is also clear from the figure
 that the peak is asymmetric due to the opening of the $f_0 K$ and $a_0 K$ thresholds involved. 
 Taking into account the relative sizes  of the peaks in the left and right panels of
 Fig.~\ref{fig:t2.k12} one infers that the $K(1460)$ couples more strongly to $f_0 K$ 
 than to  $a_0 K$, with the ratio of couplings as $|g_{f_0K}/g_{a_0K}|\simeq (\frac{20.2}{5.7})^{1/4}\simeq 1.4$~.

\begin{figure}[h]
\includegraphics[width=\textwidth,keepaspectratio]{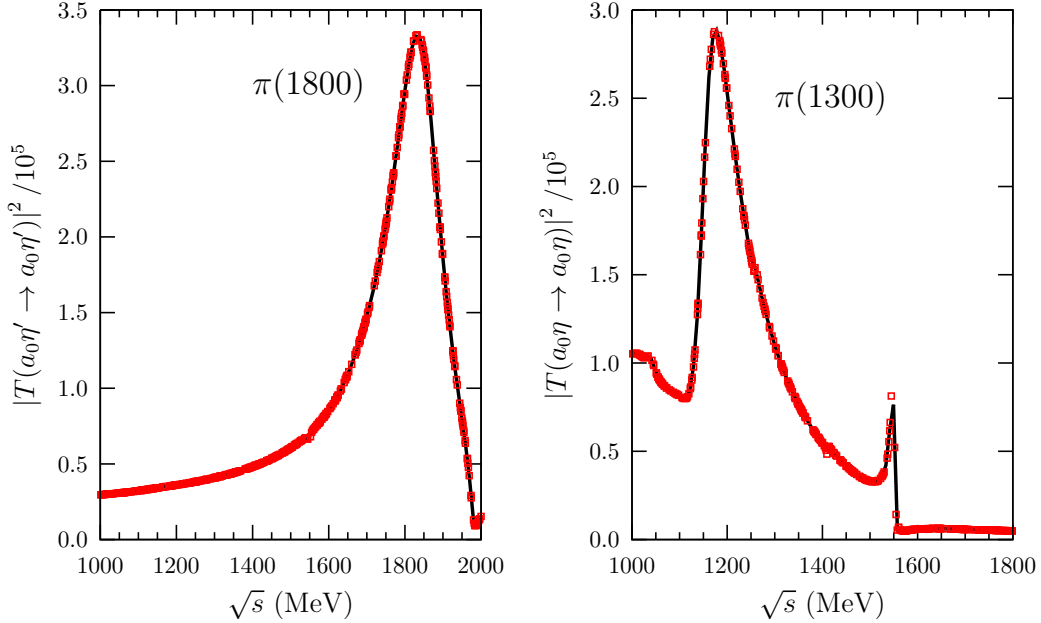}
\caption{Modulus squared of the $a_0 \eta'\to a_0 \eta'$ (left)  and 
$a_0 \eta\to a_0 \eta$ (right) S-wave amplitudes. 
 For the former $a_1=-1.3$ and for the latter $a_1=-2.0$, see the 
 text for details. The notation is as in Fig.~\ref{fig:t2.k12}.
\label{fig:t2.pi1}}
\end{figure}

\subsection{$I=1$}
We now consider the $I=1$ case. As commented in the introduction two broad resonances are referred in the PDG, the $\pi(1300)$ and $\pi(1800)$. 
In our amplitudes we find quite independently of the value of $a_1$ that 
the $a_0(980)\eta'$ channel is almost elastic. This is due to the fact that 
the interaction kernels ${\cal T}(a_0\eta'\to a_0\eta)$ and ${\cal T}(a_0\eta'\to f_0\pi)$ are much smaller than the rest of kernels, typically by an order of magnitude. This happens because the kernels are dominated by the threshold region. However, the threshold for $a_0(980)\eta'$  is much 
higher than the thresholds for the other two channels. In this way, 
for the inelastic processes involving the $a_0\eta'$ channel, even at threshold 
for one of the channels, there is always a large three-momentum for the
 other channel and the kernel is suppressed. Of course, this does not apply for the $a_0\eta'$ elastic case where the kernel has a standard size and produces around 1.8~GeV a strong resonant signal that could be associated with the 
 $\pi(1800)$ resonance. To reproduce the mass value given in the PDG \cite{pdg} for this resonance, $1816\pm 14$~MeV one takes $a_1$ for $a_0 \eta'$ around -1.3. The visual width of the peak is around 200~MeV, close to the width quoted in the 
 PDG \cite{pdg} of $208\pm 12$~MeV. The other two channels couple 
 quite strongly between each other and typically give rise to an enhancement between 1.2-1.4~GeV when varying $a_1$ equal for each of them, which could be associated with the $\pi(1300)$. However, 
 for $|a_1|$ between 1 and  1.8 a too strong signal in the $a_0\eta$ threshold originates. For $|a_1|$ below 1 the resonant peak in the $|T(a_0\eta\to a_0\eta)|^2$  lies around 1.4-1.5~GeV, somewhat too high for the $\pi(1300)$ resonance \cite{pdg}. This is why we show in Fig.~\ref{fig:t2.pi1} the modulus squared 
 of $a_0\eta\to a_0\eta$ for $a_1=-2$ where a peak close to 1.2~GeV is seen with a width of around 200~MeV. One can also see the strong effect of the $a_0 \eta$ threshold at around 1.52~GeV. Its  size is rather sensitive
 to the actual vale of $|a_1|$ when this lies between 1 and 1.8. 
 There is the interesting fact, which is independent of the value of $a_1$, 
 that there is no signal for $\pi(1800)$ in the $a_0\eta$ system nor 
 signal of the peak at 1.2~GeV in the $a_0\eta'$. We have also 
 checked that this is also the case 
 for the $f_0\pi$ state, that is, it does not couple with the $\pi(1800)$. This is another reflection of the fact that the 
  $a_0\eta'$ tends to decouple from the other states.

\begin{figure}[h]\centering
\includegraphics[width=\textwidth,keepaspectratio]{I0.eps}
\caption{Modulus squared of the $f_0 \eta'\to f_0 \eta'$ (left)  and 
$f_0 \eta\to f_0 \eta$ (right) S-wave amplitudes. 
 For the former $a_1=-1.25$ and for the latter $a_1=-0.8$, see the 
 text for details. The notation is as in Fig.~\ref{fig:t2.k12}.
\label{fig:t2.ei0}}
\end{figure}

\subsection{$I=0$}
We move next to the $I=0$ system where the $f_0\eta$, $a_0\pi$ and $f_0\eta'$ couple. Here occurs similarly to  $I=1$, so that the much higher $f_0\eta'$ channel mostly decouples from the other two channels. We then proceed similarly and distinguish between the subtraction constant $a_1$ attached to $a_0\eta'$ and to the other two channels $a_0\pi$ and $f_0\eta$. For $a_1$ around $-1.2$ one obtains a resonance of the $a_0\eta'$ channel at a mass of 1835~MeV, in agreement with that quoted in the PDG for the $X(1835)$, $1833.7\pm 6.1\pm 2.7$~MeV. This is shown in the left panel of Fig.~\ref{fig:t2.ei0} where 
the modulus squared of the $f_0(980)\eta'\to f_0(980) \eta'$ 
S-wave amplitude is shown. 
The width of the peak at half its maximum value is around 70~MeV, in good agreement with the width given in the PDG for the $X(1835)$ of $67.7\pm 20.3\pm 7.7$~MeV.   We consider next the other two coupled channels, $a_0(980)\pi$ and  $f_0(980) \eta$. We obtain a clear resonant signal with mass around 1.45~GeV for $|a_1|\lesssim 1$. This is shown in the right panel of Fig.~\ref{fig:t2.ei0}, where the modulus squared of the $f_0(980)\eta\to f_0(980)\eta$ is given for $a_1=-0.8$. It is not possible to increase further the mass of this peak by varying $a_1$. An important fact of this resonance is that it does not couple to the $a_0\pi$ channel. E.g. the analogous curve for the modulus squared of the $a_0(980)\pi\to a_0(980)\pi$ S-wave in the 1.4~GeV region is absolutely flat. By considering the inelastic process $f_0(980)\eta\to a_0(980)\pi$ we estimate a coupling to the latter channel more than 14 times smaller than to $f_0(980)\eta$. Because the $\eta(1405)$ resonance couples 
mostly to $a_0(980)\pi$ \cite{pdg} we then conclude that the generated resonant signal 
around $1.45$~GeV should correspond to the $\eta(1475)$. Its form is rather asymmetric due to the opening of the $f_0(980)\eta$ threshold, with a width at half the maximum of its peak of around 150~MeV. The width quoted in the PDG \cite{pdg} is $85\pm 9$~MeV. It is also known that the $\eta(1475)$ couples strongly to $K^*(892)\bar{K}+c.c$, a channel not included in our study. The threshold for 
this channel, at around 1.39~GeV at the decreasing slop of our present signal, should certainly modify its shape.     
For higher values of $|a_1|$ the peak tends to become too light in mass 
compared with the $\eta(1475)$. For the $a_0(980)\pi\to a_0(980)\pi$ reaction one also appreciates a strong $a_0(980)\pi$ threshold effect at around 1.16~GeV. No resonance  around the mass of the $\eta(1295)$ is observed.

\begin{figure}[h]\centering
\includegraphics[width=.5\textwidth,,keepaspectratio]{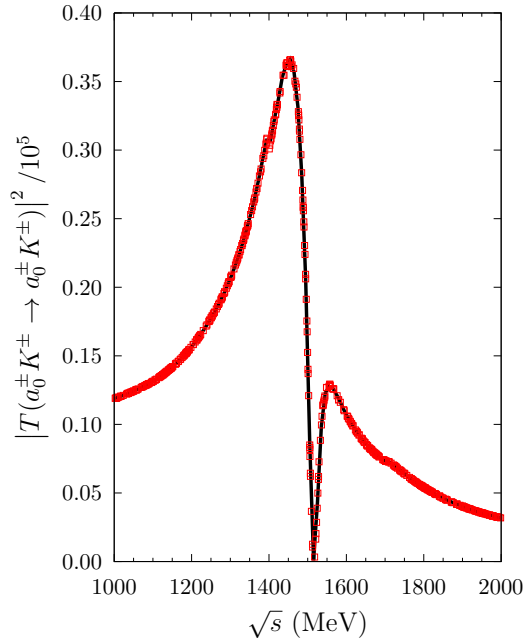}
\caption{ Modulus squared of the $I=3/2$ $a_0 K\to a_0 K$  S-wave amplitude with 
$a_1=-0.5$.  The notation is as in Fig.~\ref{fig:t2.k12}.
\label{fig:t2.ex}}
\end{figure}

\begin{table}
\begin{center}
\begin{tabular}{lccl}
\hline
Resonance & $I^{(G)}$ & Width (MeV) & Comments \\
\hline
$K(1460)$ & $I=\frac{1}{2}$ & $\Gamma\gtrsim 100$ & $|g_{f_0 K}/g_{a_0 K}|\simeq 1.4$ \\
$\pi(1800)$ & $I^G=1^{-}$ & $\Gamma\simeq 200$  & $a_0 \eta'$ elastic \\
$\pi(1300)$ & $I^G=1^{-}$ & $\Gamma\gtrsim 200$ & $a_0\pi$, $f_0\eta$ coupled channels \\
$X(1835)$ & $I^G=0^+$ & $\Gamma\simeq 70$       & $f_0\eta'$ elastic \\
$\eta(1475)$ & $I^G=0^+$ & $\Gamma\simeq 150$   & $f_0\eta$ elastic\\
Exotic & $I=\frac{3}{2}$ & $\Gamma\simeq 200$   & $a_0 K$ threshold \\
\hline
\end{tabular}
\caption{\protect \small Resonances resulting from our study. For more details see the discussions of the results in the text.
\label{tab:res}}
\end{center}
\end{table}

\subsection{Exotic channels}
Regarding the exotic channel with $I=3/2$ we find an interesting result. Our  amplitude gives rise to a clear resonant structure at around 1.4~GeV for  $|a_1|\lesssim 1.5$. We show  the modulus squared of $T(a_0^{\pm}K^\pm\to a_0^\pm K^\pm)$, because the $a_0^\pm K^\pm$ states are purely $I=3/2$,  for $a_1=-0.5$ (the same value used before in Fig.~\ref{fig:t2.k12} studying the $I=1/2$ case)  in  Fig.~\ref{fig:t2.ex}. One also observes that the shape of the resonance peak is 
asymmetric with a clear impact of the $a_0 K$ threshold. 
 Our results for $|a_1|\lesssim 1$ tends to confirm the predictions of  Longacre \cite{longacre} that studied the $K\bar{K}\pi$ and $K\bar{K}K$ system and concluded that the exotic $I=3/2$ $J^{P}=0^{-}$  $K \bar{K}K$ system was resonant around its threshold due to the successive interactions between a $K$, $\bar{K}$ and a $\pi$.  For $|a_1|\gtrsim 1$ we find that the resonance shape in $|T(a_0^\pm K^\pm\to a_0^\pm K^\pm)|^2$ progressively distorts becoming lighter and flatter.  Let us notice also that the $a_0 K$ system was not isolated in the two experiments quoted in the PDG where the $I=1/2$ $K(1460)$ was observed \cite{daum,branden}.

The other exotic channel with $I=1$ and $G=+1$ involves the isovector $a_0\pi$ state. Whether a resonance behavior stems at around 1.4~GeV depends on the actual value of $a_1$. For $|a_1|\lesssim 1$ the enhancement near 1.4~GeV is much weaker and is overcome by the cusp effect at the $a_0\pi$ threshold.
 For larger values of $|a_1|$ the resonant signal is much more prominent. 
 No such resonance has  been found experimentally, e.g. in peripheral hadron production \cite{kek}, so that $|a_1|\lesssim 1$ should be finally taken.   

In Table~\ref{tab:res} we collect all the resonances found in our study for the different quantum numbers discussed.

\section{Summary and conclusions}
\label{sec4}

In summary, we have presented a study of the S-wave interactions between the scalar resonances $f_0(980)$ and $a_0(980)$ with the lightest pseudoscalars ($\pi$, $K$, $\eta$ and $\eta'$) in the region between 1 and 2~GeV. The different channels studied comprise those alike the $\eta$, $K$ and $\pi$,   and the exotic ones with isospin 3/2 and 1, the latter having positive G-parity.
 First,  interaction kernels have been derived by considering the interactions of the external pseudoscalars involved in the reaction with those making the scalar resonance. We take advantage here of previous studies that establish  the $f_0(980)$ and 
$a_0(980)$ as dynamically generated from the interactions of two pseudoscalars,   so that no free parameters are introduced in their calculation. Afterwards, the final S-wave amplitudes are determined by employing techniques 
borrowed from Unitary Chiral Perturbation Theory. Interestingly, we have obtained resonant peaks that for the non-exotic channels could be associated with the pseudoscalar resonances $K(1460)$, $\pi(1300)$, $\pi(1800)$, $\eta(1475)$ and 
$X(1835)$, following the notation of the particle data group. 
 The resonances that come out from this study can be qualified as dynamically generated from the interactions between the scalar resonances and the pseudoscalar mesons. This establishes that an important contribution to the physical signal of the resonances just mentioned has a dynamical origin. 
  The exotic $I=3/2$ channel could also exhibit a resonant structure around the $a_0 K$ threshold, in agreement with the behavior predicted 
by Longacre \cite{longacre} twenty years ago. However, larger values for the subtraction constant 
 $|a_1|$ tends to destroy this resonant behavior. No signal of the intriguing $\eta(1405)$ resonance is obtained. 

 This approach should be pursued further by including simultaneously 
 to the interaction between the 
scalar resonances and the pseudoscalar mesons, considered here, those 
arising from the lightest vector resonances with the same pseudoscalars in 
P-wave. In this way, both pseudoscalar and axial resonances will be studied together.

\section*{Acknowledgements}
This work has been partially funded by the MEC grant FPA2007-6277 and Fundaci\'on S\'eneca grant 11871/PI/09. M.A. acknowledges the Fundaci\'on S\'eneca for the FPI grant 13310/FPI/09. We also acknowledge the financial support from  the BMBF grant 06BN411, EU-Research Infrastructure Integrating Activity "Study of Strongly Interacting Matter" (HadronPhysics2, grant n. 227431)
under the Seventh Framework Program of EU and   the Consolider-Ingenio 2010 Program CPAN (CSD2007-00042). Computing time and support was provided by Parque Cient\'{\i}fico de Murcia in the Ben Arabi SuperComputer.

\appendix{}

\section{Meson--meson unitarized amplitudes}
\label{app:fits}
\def\theequation{\Alph{section}.\arabic{equation}}
\setcounter{equation}{0}

We use the $N/D$ method \cite{nd}
 to unitarize the different isospin channels amplitudes for meson--meson scattering, which are fitted to data, 
and then used in the vertex of the triangle loop. From these amplitudes, once fitted, the position of 
the poles can be found (we use here the $f_0(980)$ and $a_0(980)$, but we also check for 
the appearance of the other scalars, $\sigma$ and $\kappa$). As mentioned, the amplitudes are unitarized through
\begin{equation}
T_{I} = (1+ V_{I}\cdot G)^{-1}\cdot V_{I}~,
\end{equation}
which is analogous to eq.~\eqref{t.uni} but now for the pseudoscalar-pseudoscalar scattering.  
The symmetric matrix $V_{I}$ (the analogous one to ${\cal T}_{IG}$ in eq.~\eqref{t.uni}) collects the 
S-wave pseudoscalar-pseudoscalar tree-level amplitudes  obtained from the lowest order Chiral Lagrangians  including resonances as well. 
The matrix $G$ is a diagonal matrix that contains the meson--meson loop propagator (the same expression as 
given in eq.~\eqref{g.func} can be used with the appropriate replacement for the masses involved.)

The lowest order chiral Lagrangian at leading order in large $N_c$ which also includes the $\eta_1$  is \cite{wit,barna,roland}
\begin{eqnarray}
\L_2 & = & \frac{f_\pi^2}{4}\langle \partial_{\mu}U^\dagger \partial^{\mu}U \rangle + \frac{f_\pi^2}{4}\langle \chi^{\dagger}U + \chi U^{\dagger} \rangle - \frac{1}{2}M_1^2 \eta_1^2 \label{lag-p2}~,\\
U(\phi) & = & \exp\left(i \sqrt{2}\Phi/f_\pi \right)\text{,} \quad 
\Phi = \sum_{i=0}^{8} \frac{\lambda_i}{\sqrt{2}}\phi_i\text{,} \quad \lambda_0 = \sqrt{\frac{2}{3}}\mathbf{I}_3~,
\label{l2}
\end{eqnarray}
and $\lambda_i$,~$i=1,\ldots,8$ the Gell-Mann matrices. 
The $\eta_8$ and $\eta_1$ fields mix to give the physical $\eta$ and $\eta'$
\begin{equation*}
\left( \begin{array}{c}\eta' \\ \eta \end{array} \right) = \left( \begin{array}{cc} \cos{\theta} & \sin{\theta} \\ -\sin{\theta} & \cos{\theta} \end{array}\right) \left( \begin{array}{c}\eta_1 \\ \eta_8 \end{array} \right)~.
\end{equation*}
The mixing angle $\theta$ is taken as $\sin\theta \simeq -\frac{1}{3}$, $\theta \simeq -20^\circ$ \cite{jamin}.

\begin{figure}\centering
\includegraphics[height=2cm,keepaspectratio]{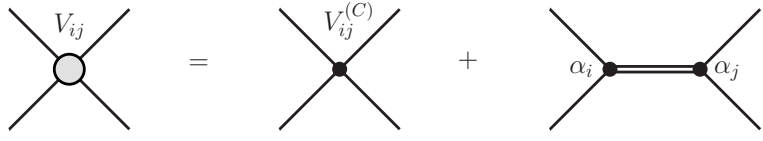}
\caption{Eq.~\eqref{eq:contplusreson} in terms of Feynman Diagrams\label{fig:contplusreson}}
\end{figure}

In the same spirit as in ref.~\cite{nd} the explicit exchange of $J^{PC}=0^{++}$ scalar resonances is incorporated and calculated 
from the leading order chiral Lagrangians of ref.~\cite{Ecker:1988te}. The appropriate Lagrangians are
\begin{eqnarray}
\L_{S_8} & = & c_d \left\langle S_8 u_{\mu}u^{\mu} \right\rangle + c_m \left\langle S_8 \chi_+ \right\rangle~, \label{lag-res-oct}\\
\L_{S_1} & = & \tilde{c}_d  S_1 \left\langle u_{\mu}u^{\mu} \right\rangle + \tilde{c}_m S_1 \left\langle \chi_+ \right\rangle~, \label{lag-res-sin}\nn \\
\chi_{+} & = & u^\dagger \chi u^\dagger + u \chi^{\dagger} u~, \nn \\
U(x) & = & u(x)^2 \text{,}\quad u_{\mu} = i u^{\dagger}\partial_{\mu} U u^{\dagger} = u_{\mu}^{\dagger}~, \nn \\
S_8 & = & \left(
\begin{array}{ccc}
\frac{a_0}{\sqrt{2}}+\frac{f_8}{\sqrt{6}} & a_0^+ & {K^*_0}^+\\
a_0^- & -\frac{a_0}{\sqrt{2}}+\frac{f_8}{\sqrt{6}} & {K^*_0}^0\\
{K^*_0}^- & {\overline{K}^*_0}^0 & -\frac{2}{\sqrt{6}}f_8
\end{array}
\right)~, \nn
\label{lag}
\end{eqnarray}
and $S_1$ is a scalar SU(3) singlet.
The interaction kernels obtained from Lagrangians \eqref{lag-p2}, \eqref{lag-res-oct} and \eqref{lag-res-sin} can thus be written as
\begin{eqnarray}
V_{ij} & = & V^{(C)}_{ij} + V^{(R)}_{ij}~,\label{eq:contplusreson} \\
V_{ij}^{(R)} &= & \frac{\alpha_i \alpha_j}{M_R^2-s}~, \nn
\end{eqnarray}
where $C$ means \textit{contact term} and $R$ \textit{resonance exchange}. 
This is represented  diagrammatically in Fig.~\ref{fig:contplusreson}.
In what follows, we give explicit formulae for the contact kernels
 from the chiral Lagrangians in eq.~\eqref{l2} for the different isospin channels. 
 We also give the couplings $\alpha_i$ for each scalar resonance. For $I=0$ we include the superscript $(8)$ or 
$(1)$ in $\alpha_i$ to distinguish between the octet and singlet contributions, respectively. For the 
rest of isospins there is no singlet contribution.

\begin{itemize}
\item{$I=0$}
\begin{eqnarray}
V_{\pi\pi \to \pi\pi}^{(C)}     & = &  \frac{2s-m_{\pi}^2}{2f_{\pi}^2} \nn\\
V_{\pi\pi \to K\bar{K}}^{(C)}   & = &  \frac{\sqrt{3}}{4} \frac{s}{f_{\pi}^2} \nn\\
V_{\pi\pi \to \eta\eta}^{(C)}   & = & -\frac{1}{\sqrt{3}}\frac{m_{\pi}^2}{f_{\pi}^2} \nn\\
V_{K\bar{K} \to K\bar{K}}^{(C)} & = & \frac{3}{4}\frac{s}{f_{\pi}^2} \nn\\
V_{K\bar{K} \to \eta\eta}^{(C)} & = & -\frac{2}{9}\frac{3s-2 m_{\eta}^2-m_{K}^2}{f_{\pi}^2} \nn\\
V_{\eta\eta \to \eta\eta}^{(C)} & = &  \frac{2m_{K}^2+m_{\pi}^2}{9f_{\pi}^2}
\end{eqnarray}

\begin{eqnarray}
\alpha^{(8)}_{\pi\pi}   & = &  \frac{c_d s + 2(c_m - c_d)m_\pi^2}{f_\pi^2} \nn\\
\alpha^{(8)}_{K\bar{K}} & = & -\frac{c_d s + 2(c_m - c_d)m_K^2}{\sqrt{3}f_\pi^2} \nn\\
\alpha^{(8)}_{\eta\eta} & = &  \frac{8 c_m (m_{K}^2-m_\pi^2)}{3\sqrt{3}f_\pi^2}
\end{eqnarray}
\begin{eqnarray}
\alpha^{(1)}_{\pi\pi}   & = &  \sqrt{6}\frac{\tilde{c}_d s + 2(\tilde{c}_m - \tilde{c}_d)m_\pi^2}{f_\pi^2} \nn\\
\alpha^{(1)}_{K\bar{K}} & = & -\frac{\tilde{c}_d s + 2(\tilde{c}_m - \tilde{c}_d)m_K^2}{2\sqrt{2}f_\pi^2} \nn\\
\alpha^{(1)}_{\eta\eta} & = &  \sqrt{2}\frac{3 \tilde{c}_d (s-2m_{\eta}^2) + 2 \tilde{c}_m (2m_{K}^2+m_\pi^2)}{3 f_\pi^2}
\end{eqnarray}

\item{$I=1/2$}
\begin{eqnarray}
V_{K\pi \to K\pi}^{(C)}     & = &  \frac{s-3u+2m_{K}^2+2m_{\pi}^2}{4f_{\pi}^2} \nn\\
V_{K\pi \to K\eta}^{(C)}    & = &  \frac{\sqrt{2}}{6}\frac{3t-m_{\eta}^2-2m_{K}^2}{f_{\pi}^2} \nn\\
V_{K\pi \to K\eta'}^{(C)}   & = & -\frac{1}{12}\frac{3t-8m_{K}^2-m_{\eta'}^2-3m_{\pi}^2}{f_{\pi}^2} \nn\\
V_{K\eta \to K\eta}^{(C)}   & = & \frac{6t-4m_{\eta}^2-2m_{K}^2}{9f_{\pi}^2} \nn\\
V_{K\eta \to K\eta'}^{(C)}  & = & -\frac{1}{9\sqrt{2}}\frac{3t - m_{\eta}^2-m_{\eta'}^2-3m_{\pi}^2+2m_{K}^2}{f_{\pi}^2} \nn\\
V_{K\eta' \to K\eta'}^{(C)} & = &  \frac{1}{36}\frac{3t - 2m_{\eta'}^2+32m_{K}^2-6m_{\pi}^2}{f_{\pi}^2}
\end{eqnarray}
\begin{eqnarray}
\alpha_{K\pi}   & = & -\sqrt{\frac{3}{2}} \frac{c_d s + (c_m - c_d)(m_\pi^2 + m_K^2)}{f_\pi^2} \nn\\
\alpha_{K\eta}  & = & -\frac{2}{\sqrt{3}} \frac{c_m (m_K^2 - m_\pi^2)}{f_\pi^2} \nn\\
\alpha_{K\eta'} & = & -\sqrt{\frac{3}{2}} \frac{c_d s -c_d(m_K^2 + m_{\eta'}^2)
 - \frac{1}{3}c_m(m_\pi^2 - 7m_{K}^2) }{f_\pi^2}
\end{eqnarray}

\item{$I=1$}
\begin{eqnarray}
V_{\pi\eta  \to \pi\eta}^{(C)}   & = &  \frac{2}{3}\frac{m_{\pi}^2}{f_{\pi}^2} \nn\\
V_{\pi\eta  \to K\bar{K}}^{(C)}  & = &  -\frac{3s-2m_{K}^2-m_{\eta}^2}{3\sqrt{3}f_{\pi}^2} \nn\\
V_{\pi\eta  \to \pi\eta'}^{(C)}  & = &  \frac{\sqrt{2}}{3}\frac{m_{\pi}^2}{f_{\pi}^2} \nn\\
V_{K\bar{K} \to K\bar{K}}^{(C)}   & = & -\frac{u-2m_{\pi}^2}{2f_{\pi}^2} \nn\\
V_{K\bar{K} \to \pi\eta'}^{(C)}  & = &  \frac{3s-8m_{K}^2-m_{\eta'}^2-3 m_{\pi}^2}{6\sqrt{6}f_{\pi}^2} \nn\\
V_{K\eta' \to K\eta'}^{(C)} & = &  \frac{1}{3}\frac{m_{\pi}^2}{f_{\pi}^2}
\end{eqnarray}
\begin{eqnarray}
\alpha_{\pi\eta}  & = & -\frac{2}{\sqrt{3}} \frac{c_d s - c_d(m_{\pi}^2+m_{\eta}^2)  + 2 c_m m_{\pi}^2}{f_\pi^2} \nn\\
\alpha_{K\bar{K}} & = &                    \frac{c_d s - 2 (c_d -c_m) m_K^2}{f_\pi^2} \nn\\
\alpha_{\pi\eta'} & = & -\frac{2}{\sqrt{6}} \frac{c_d s - c_d(m_{\pi}^2+m_{\eta'}^2) + 2 c_m m_{\pi}^2}{f_\pi^2} \nn\\
\end{eqnarray}
\item{$I=3/2$}
\begin{equation}
V_{K\pi \to K\pi} = -\frac{s-m_{K}^2-m_{\pi}^2}{2f_\pi^2}
\end{equation}
\end{itemize}

\begin{figure}[ht]\centering
\includegraphics[width=\textwidth,keepaspectratio]{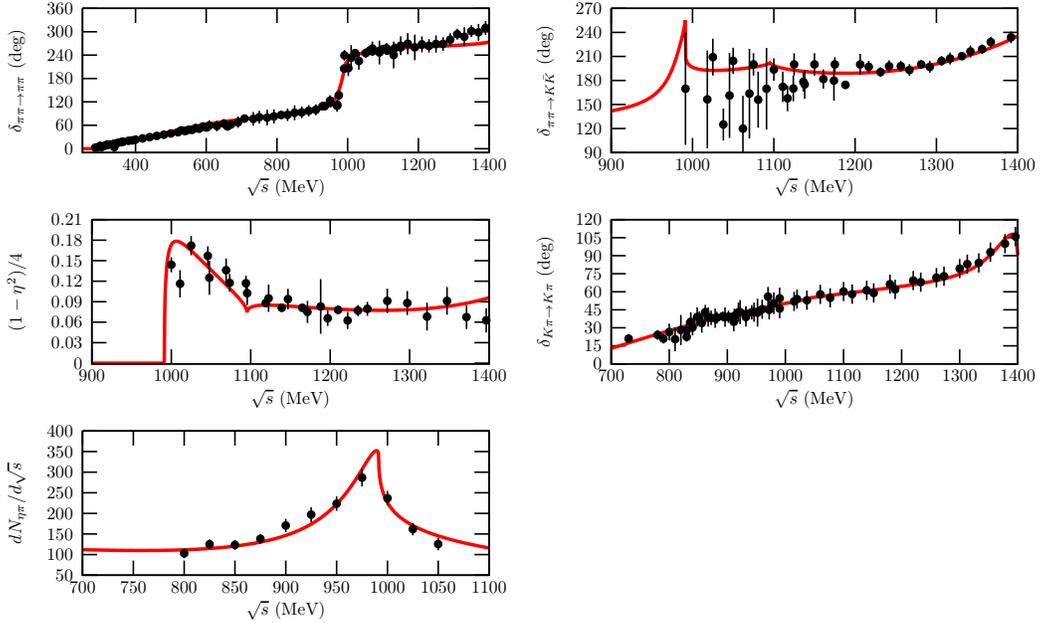}
\caption{Experimental data and our fits, as explained in the text.\label{fig:observables}}
\end{figure}

With the amplitudes calculated for meson--meson scattering, we perform several fits, e.g. by changing the value of the highest $\sqrt{s}$ fitted from 1.2 to 1.4~GeV and by imposing that several subtraction constants for the 
pseudoscalar--pseudoscalar channels are equal, so that we can calculate the 
pseudoscalar--scalar kernels with different inputs, and then check the independence of our results. 
We only show our main fit since all the other fits that we obtained 
give rise to similar results that would not change our conclusions. 
In this fit the highest value of $\sqrt{s}$ considered  is 1.4~GeV.  
For the octet of scalar resonances we take the values of the parameters 
 $c_d$, $c_m$ and $M_8$  from ref.~\cite{jamin}, where $M_8$, the mass 
 of this octet, is around 1.3-1.4~GeV. For definiteness, 
 $c_d=c_m=22.8$~MeV and $M_8=1.4$~GeV. 
   The parameters for the singlet resonance exchange, $\widetilde{c}_d$, 
   $\widetilde{c}_m$ and $M_1$ are left free, with the latter the mass 
   of the singlet scalar resonance. 
  Regarding the subtraction  constants in the unitarity loop function of the different channels \cite{nd} (they play the analogous role of $a_1$ in 
  eq.~\eqref{g.func} but for pseudoscalar-pseudoscalar scattering), we take the 
  most general situation compatible with isospin symmetry. 
  Adopting the same argument as in the appendix~A of ref.~\cite{jido} from SU(3) to SU(2), the subtraction constants corresponding to the same pair of 
  pseudoscalars should be the same in the different isospin. In this way, the subtraction constant for $K\bar{K}$ both in $I=0$ and $I=1$ is taken with 
  the same value. 
  On the other hand, for a given isospin, we also put constraints on the subtraction 
  constants associated with non-relevant channels. In this way, the 
   $K\eta$ subtraction constant in $I=1/2$ is kept equal to that of $K\pi$ and, 
   similarly, for $I=1$ the $\pi\eta'$ subtraction constant is put equal to that of $\pi\eta$.\footnote{We have checked that the $\pi\eta'$ channel tends to decouple of the $\pi\eta$ and $K\bar{K}$ channels in $I=1$ in the region of the $a_0(980)$.}  Of course, we have checked that smoothing these constraints does not affect the results of the fit.  In this way, we finally have six independent subtraction 
  constants for $\pi\pi$,  $K\bar{K}$, $\eta\eta$, $K\pi$, $K\eta'$ 
  and $\pi\eta$.  There is also a normalization constant for the data on an 
 unnormalized  $\pi\eta$ event distribution around the $a_0(980)$ resonance that is required for each fit. The results of the fit compared to experimental data 
 are shown by the solid line in Fig.~\ref{fig:observables} and the values of the fitted parameters are given in Table~\ref{tab:fit_params}.
 
\begin{table}\label{tab:fits}
\footnotesize
\centering
\begin{tabular}{cc}
Parameter & Value \\ \hline\hline
$\tilde{c}_d$ (MeV) &  $18 \pm 1$    \\
$\tilde{c}_m$ (MeV) &  $23 \pm 4$    \\
$M_1$ (MeV)         &  $1100 \pm 20$ \\ \hline
$a_{\pi\pi}$   & $-0.98 \pm 0.10$  \\
$a_{K\bar{K}}$ & $-1.00 \pm 0.20$ \\
$a_{\eta\eta}$ & $+0.04 \pm 0.22$ \\
$a_{K\pi}$   & $+0.17 \pm 0.05$ \\
$a_{K\eta'}$ & $-3.53 \pm 0.13$\\
$a_{\pi\eta}$& $-2.55 \pm 0.37$\\ \hline 
\end{tabular}
\caption{Fitted parameters for the main fit. The $\chi^2/\textrm{d.o.f.}$ is $0.96$. The fits are obtained employing the program MINUIT \cite{minuit}.
\label{tab:fit_params}}
\end{table}

\begin{table}
\footnotesize
\centering
\begin{tabular}{crr}
Resonance  & $\textrm{Re}\sqrt{s}\textrm{ (MeV)}$ & $\textrm{Im}\sqrt{s}\textrm{ (MeV)}$ \\ \hline\hline
$\sigma$   &  $466$ & $235$ \\
$\kappa$   &  $698$ & $294$ \\
$f_0(980)$ &  $987$ & $ 18$ \\
$a_0(980)$ & $1019$ & $ 33$ \\ \hline
\end{tabular}
\caption{The pole positions of the scalar resonances obtained from the main fit are given.\label{tab:fit_poles_1}}
\end{table}

\begin{table}
\footnotesize
\centering
\begin{tabular}{ccc} 
 Resonance & $g_{K\bar{K}}\textrm{ (GeV)}$ & $\left\lvert g_{K\bar{K}}\textrm{ (GeV)} \right\rvert$ \\ \hline \hline
 $f_0(980)$ & $-3.72 + 1.18i$ & $3.90$  \\ 
 $a_0(980)$ & $-4.11 + 1.59i$ & $4.41$  \\ \hline
\end{tabular}
\caption{Couplings of the  $f_0(980)$ and $a_0(980)$ resonances to $K\bar{K}$ (with definite isospin). These couplings are calculated 
from the residues of the corresponding pole, see e.g.\cite{nd,penni}.\label{tab:fit_poles_2}}
\end{table}

The set of experimental data included in the fits for $I=0$ comprises the elastic $\pi\pi$ phase shifts, 
$\delta_{\pi\pi\to\pi\pi}$, from 
refs.~\cite{Grayer:1972,Estabrooks:1973,Kaminski:1996da,Froggatt:1977hu,Hyams:1973zf,Protopopescu:1973sh}, 
the phase shift for $\pi \pi\to K\bar{K}$, $\delta_{\pi\pi\to K\bar{K}}$, and
 $(1-\eta^2)/4$ from refs.~\cite{Cohen:1980cq,Etkin:1981sg}, where $\eta$ is the elastic parameter for the 
 $\pi\pi\to \pi\pi$ $I=0$ S-wave.  With respect to $I=1/2$ we fit the elastic $\pi K$ phase shifts, $\delta_{K\pi \to K\pi}$,  
from refs.~\cite{Mercer:1971kn,Estabrooks:1977xe,Bingham:1972vy,Aston:1987ir}. Finally, we include 
an event distribution of $\pi\eta$ around the $a_0(980)$ resonance mass from the central production of $\pi\pi\eta$,  ref.~\cite{Armstrong:1991rg}, 
fitted like in ref.~\cite{iamprl,iamcc}.

Once the fits are performed we look for the poles of the scalar resonances $\sigma$, $\kappa$, $f_0(980)$ and $a_0(980)$ 
in the unphysical Riemann sheets continuously
 connected with the physical one. Notice that only  the $f_0(980)$ and $a_0(980)$ poles are  actually required  
 for evaluating the pseudoscalar--scalar scattering kernels in section~\ref{sec2}. The $\sigma$ and $\kappa$ are given for completeness. 
 They are related to the $f_0(980)$ and $a_0(980)$ resonances giving rise to a nonet of light scalar resonances \cite{mixing}. 
 Other poles around 1.4~GeV also appear that we do not include here.  
The pole positions are given in Table~\ref{tab:fit_poles_1}. 
The couplings of the $f_0(980)$ and $a_0(980)$ to $K\bar{K}$, used in this work, are collected in Table~\ref{tab:fit_poles_2}. 

\section{S-wave projection of $C_3$ and $C_4(M_4^2)$}
\label{app:c3.c4.pro}
\def\theequation{\Alph{section}.\arabic{equation}}
\setcounter{equation}{0}

 The three- and four-point Green functions $C_3$ and $C_4(m_4^2)$ are defined 
 in eq.~\eqref{c3.c4.def}. Here, we consider the more general case with arbitrary internal masses and from the very beginning
the S-wave projection is worked out. Both functions are finite. 
 
\begin{figure}[h!t]
\centering
\includegraphics[height=4.5cm,keepaspectratio]{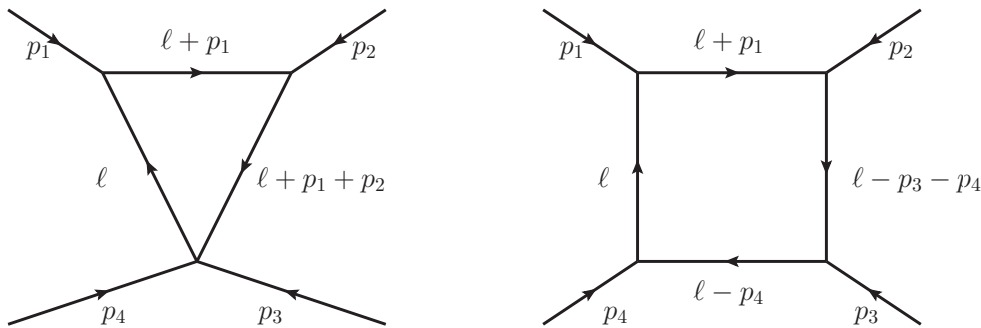}
\caption{Feynman diagrams for $C_3$ (left) and $C_4(m_4^2)$ (right).
\label{fig:c3.c4}}
\end{figure}

 We first consider $C_3$, left diagram of Fig.~\ref{fig:c3.c4}, and follow its 
 notation  with  $t=(p_1+p_2)^2$ (note that all four-momenta are in-going). We also introduce two 
 Feynman parameters $u_1$ and $u_2$ and the relative angle $\theta$ between the initial and final pseudoscalars, so that
 \begin{align}
 C_3 & =\frac{i}{2}\int_{-1}^{+1}d \cos\theta\int\frac{d^4\ell}{(2\pi)^4}\frac{1}{((\ell+p_1)^2-m_1^2)(\ell^2-m_2^2)((\ell+p_1+p_2)^2-m_3^2)}\nn\\
&=\frac{1}{32\pi^2}\int_0^1du_1 \int_0^{u_1} du_2 \int_{-1}^{+1} d \cos\theta
\left[
p_2^2 u_1^2+p_1^2 u_2^2+2 p_1 p_2 u_1 u_2\right. \nn\\ 
&\left.+(m_1^2-m_3^2-p_2^2)u_1 +(m_2^2-m_1^2-p_1^2-2 p_1 p_2)u_2+m_3^2+i\ep\right]^{-1}~.
 \end{align}
Then,  $\cos\theta$ is introduced by taking into account that $p_1 p_2=(t-p_1^2-p_2^2)/2$ with 
$t=q_0^2-|\vp|^2-|\vp'|^2+2|\vp||\vp'|\cos\theta$, with $q^0=p_1^0-p_2^0$, the difference of energies
 between the initial and final scalar resonances. 
We perform the angular integration and introduce the parameter $\xi_2$ as $u_2=u_1 \xi_2$, so that
\begin{align}
C_3 &=\frac{1}{64\pi^2|\vp||\vp'|}\int_0^1\frac{du_1}{1-u_1}\int_0^1 \frac{d\xi_2}{\xi_2}
\left[ \log(1+\psi)-\log(-1+\psi)\right]~,
\end{align}
where
\begin{align}
\psi&=\frac{1}{2|\vp||\vp'|(1-u_1)u_2}\left[p_2^2 u_1^2+p_1^2 u_2^2+u_1 u_2(q_0^2-|\vp|^2-|\vp'|^2-p_1^2-p_2^2)\right.\nn\\
&\left.
+u_1(m_1^2-m_3^2-p_2^2)+u_2(m_2^2-m_1^2+p_2^2-q_0^2+|\vp|^2+|\vp'|^2)+m_3^2-i\ep\right]^{-1}~.
\end{align}

For the four-point function $C_4(m_4^2)$, right diagram of Fig.~\ref{fig:c3.c4},  
  one has 
\begin{align}
C_4(m_4^2)&=\frac{i}{2}\int_{-1}^{+1}d\cos\theta\int \frac{d^4\ell}{(2\pi)^4}
\frac{1}{((\ell+p_1)^2-m_1^2)(\ell^2-m_2^2)}\nn\\
&\times \frac{1}{((\ell-p_3-p_4)^2-m_3^2)((\ell-p_4)^2-m_4^2)}~.
\end{align}
In this case there is no ambiguity if instead of performing the $\cos\theta$ integration one directly 
calculates the related integration over $t=(p_3+p_4)^2$ by taking into account that $dt=2|\vp||\vp'|d\cos\theta$ (ambiguities 
could arise for $s$ such that the product $|\vp||\vp'|$ becomes complex. The particular integration to be performed here 
is not affected by such  problem, see below.) 
We also introduce three Feynman parameters $u_1$, $u_2$ and $u_3$ so that
\begin{align}
C_4(m_4^2)&=\frac{-1}{64\pi^2|\vp||\vp'|} \int_0^1 du_1 \int_0^{u_1}du_2 \int_0^{u_2}du_3\int_{t_-}^{t_+} dt
\left[
p_1^2(1-u_2+u_3)(u_2-u_3)
\right.\nn\\
&\left. +2 p_1p_3 (u_1-u_2)(u_2-u_3)
+2 p_1 p_4(1-u_2)(u_2-u_3)
-m_2^2 u_3 \right.\nn\\
&\left.
+p_3^2(1-u_1+u_2)(u_1-u_2)
+2 p_3 p_4(u_1-u_2)u_2
+p_4^2(1-u_2)u_2
\right.\nn\\
&\left.-m_1^2(u_2-u_3)-m_3^2 (u_1-u_2)-m_4^2(1-u_1)-i\ep
\right]^{-1}~.
\end{align} 
In terms of the variable $t$ the previous integral is of the from $\int dt/(a t+b)^2$ so that the $t$-integration can be done 
straightforwardly without problems in its analytical extrapolation. The resulting $u_3$-integration is of the form 
$\int du_3/[u_3(u_3^2+\beta u_3+\gamma)]$ that can also be done straightforwardly by factorizing the second-order polynomial in the 
denominator. Our final expression for $C_4(m_4^2)$ is 
\begin{align}
C_4(m_4^2)=\frac{-1}{64 p_1^2|\vp||\vp'|}\int_0^1du_1\int_0^1 \frac{d\xi_2}{1-\xi_2}&\left[\psi(t_+,u_2)-\psi(t_+,0)\right.\nn\\
&\left.-\psi(t_-,u_2)+\psi(t_-,0)\right]~,
\end{align}
where
\begin{align}
\psi(t,u_3)&=\frac{\log u_3}{y_1 y_2}+\frac{\log(u_3-y_1)}{y_1(y_1-y_2)}-\frac{\log(u_3-y_2)}{y_2(y_1-y_2)}~,\nn\\
u_3^2+\beta u_3+\gamma&=(u_3-y_1)(u_3-y_2)~,\nn\\
p_1^2(u_3^2+\beta u_3+\gamma)&=m_4^2(1-u_1)+m_3^2(u_1-u_2)+p_3^2(-1+u_1)(u_1-u_2)+m_1^2 u_2\nn\\
&+
u_2(s(-1+u_1)+p_2^2(u_2-u_1))-(m_1^2-m_2^2+p_4^2)u_3+p_1^2 u_3^2\nn\\
&+(s(1-u_1)+(p_2^2+p_4^2-t)u_1
-(p_1^2+p_2^2-t)u_2)u_3
-i\ep~,\nn\\
t_{\pm}&=(p_3^0-p_4^0)^2-(|\vp|\mp |\vp'|)^2~.
\end{align}

The S-wave projection of the 
three- and four-point functions $C_3$ and $C_4(m_4^2)$ was also 
obtained for some kinematical regions and values of $m_4^2$ from ref.~\cite{one}, wherever the latter 
could be applied. In such cases our results and ref.~\cite{one} agree.

\end{document}